\documentclass[twocolumn,showpacs]{revtex4}
\usepackage{graphicx}%

\begin{document}

\title{The Effect of the Third Dimension on Rough Surfaces Formed by Sedimenting
Particles in Quasi-Two-Dimensions}

\author{K.~V. McCloud}
\affiliation{Department of Physics and Engineering, Xavier
University of Louisiana, New Orleans, LA 70125}
\author{M.~L. Kurnaz}
\affiliation{Department of Physics, Bogazici University,80815
Bebek Istanbul}

\begin{abstract}

The roughness exponent of surfaces obtained by dispersing silica
spheres into a quasi-two-dimensional cell is examined.   The cell
consists of two glass plates separated by a gap, which is
comparable in size to the diameter of the beads. Previous work has
shown that the quasi-one-dimensional surfaces formed have two
distinct roughness exponents in two well-defined length scales,
which have a crossover length about 1cm. We have studied the
effect of changing the gap between the plates to a limit of about
twice the diameter of the beads.

\end{abstract}

\maketitle

\section{INTRODUCTION\protect\\ }
\label{sec:level1}

The formation of rough surfaces is a problem of theoretical and
practical importance in many areas, ranging from the study of
fundamental nonequilibrium statistical physics to various
industrial processes such as the growth of films by deposition
\cite{Vicsek92book, Barabasi95book}.   Rough surfaces formed by
the sedimentation of particles through a viscous fluid are
particularly interesting, since interactions between the
sedimenting particles, and between the particles and the walls of
the container, can be expected to have an effect on the final
interface. Surfaces formed by sedimentation are close to the
original problem of sedimentation of particles sedimenting along
straight vertical lines first studied by Edwards and Wilkinson
\cite{Edwards82}. However, since the hydrodynamic
particle/particle and particle/wall forces are in principle
long-range, the rough surfaces formed by particles sedimenting in
a viscous fluid are a different growth situation from the simpler
vertical deposition. The situation is further complicated by the
presence of backflows of fluid caused by the motion of significant
numbers of particles \cite{Davis85,Auzerais88}.

In this work we are primarily interested studying experimentally
the effect of the particle/wall interactions on the roughness of
the final interface. The motion of a sphere parallel to a single
wall is of interest as the limiting case of motion of a small
sphere in a cylindrical container when the sphere approaches the
cylinder wall. This problem and the more general one of the motion
of a sphere parallel to two external walls were treated in the
1920s by Faxen \cite{Faxen23}. Unfortunately, it is very difficult
to comment on the exact nature of the interactions between the
particles in the presence of the walls. Analytical sedimentation
theory has succeeded only in analyzing the effective settling
velocity of particles in a dilute regime in the presence of the
walls \cite{
Smoluchowski11,Smoluchowski13,Burgers41,Brenner58,Kynch59,Hasimoto59,Faxen25}
and some features of many-body interactions between the particles
\cite{Mazur82,VanSaarlos83} when there are no walls
\cite{Happel65}. Recent theoretical
\cite{Auzerais88,Brady88B,Koch91} and experimental
\cite{Nicolai95,Xue92,Segre97} work hold out some hope of
determining particle interactions through a wide range of volume
fractions and Peclet numbers in sedimentation problems. However so
far there has been no experimental or theoretical work to
distinguish between the effects of the interaction between the
particles versus the interactions between the particles and the
walls, and to correlate these interactions with the surface growth
problem.

In our previous work \cite{Kurnaz93,Kurnaz96} on the
quasi-one-dimensional surfaces formed by particles sedimenting
through a viscous fluid in a quasi-two-dimensional cell we have
found that the surfaces formed by sedimenting particles are rough
on all length scales between the particle size and the cell size.
However, different roughness exponents were found in two
well-defined length-scale regimes, with a well-defined crossover
length scale.   These roughness exponents and the crossover length
scale have been found to be independent of the cell aspect ratio
or the viscosity of the fluid through which the particles settle.
The exponent found at long length scales has been shown to depend
on the rate at which particles are deposited into the cell (hence
to the strength of the interaction between the particles)
\cite{McCloud97}. This leads to the conclusion that the scaling
exponent seen at long length scales depends on the details of the
hydrodynamic interactions between the particles, while the
exponent seen at small length scales, which remained relatively
unaffected by changes in the deposition rate, may be due to more
universal considerations.

\begin{figure}[bp]
\includegraphics[width=8cm]{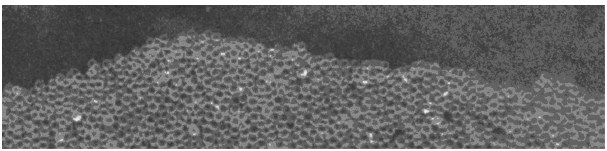}
\caption{Portion of a rough surface formed by 0.6 mm diameter
glass beads sedimenting through heavy paraffin oil.} \label{fig1}
\end{figure}

In the present work, we are continuing to study the rough surfaces
formed by sedimenting particles, but we are focusing specifically
on the effect of the particle/wall interactions on the final
interface while keeping the average volume fraction of the
particles at the low value of $\phi = 0.01$ to restrict the
particle / particle interactions to a minimum level allowed by the
experimental setup.    All of our work to date has taken place in
quasi-two-dimensional sedimentation cells, which consist of two
glass plates separated by a gap set by Teflon spacers, through
which glass beads are allowed to settle to form a quasi-one
dimensional rough surface at the bottom of the cell
(Fig.~\ref{fig1}). There are two types of cells, open and closed,
which will be discussed in more detail below.   In each type of
cell, the gap between the plates, although variable, has been of
the order of the diameter of the sedimenting particles.

\section{EXPERIMENTAL METHOD\protect\\ }
\label{sec:level2} 
\begin{figure}[bp]
\includegraphics[width=8cm]{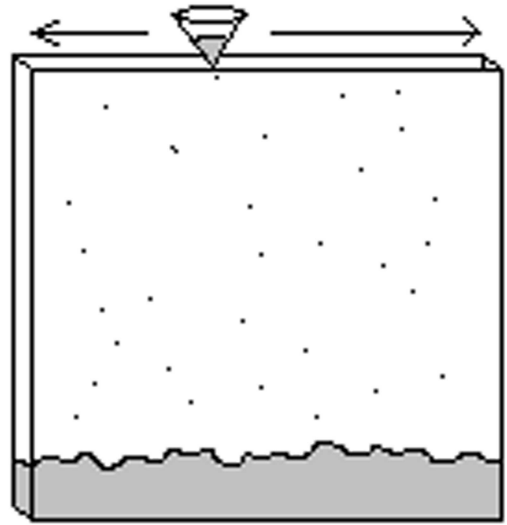}
\caption{The sedimentation cell consists of two glass plates with
a small gap between them, of the order of the diameter of the
glass beads.   The cell is filled with a viscous fluid such as
oil, and a funnel sweeps across the top of the cell, delivering a
mixture of oil and beads to the cell.   The beads settle to the
bottom of the cell and build a rough surface.} \label{fig2}
\end{figure}

Two types of cells were used in the previous
work, denoted as "closed" and "open" cells.   Closed cells were
constructed of 1/4 in. float glass, held 1mm apart by sealed side
frames of precision machined Plexiglas. Around 10,000 0.6
mm-diameter monodisperse silica spheres were placed in the cell,
which was then filled with a viscous fluid (such as glycerin) and
closed.   Each cell could be rotated about a horizontal axis
perpendicular to the gap direction.   When the cell was rotated,
the particles which had been at rest on the bottom fell though the
viscous fluid, slowly building up a new surface at the bottom of
the cell.   In the closed cells we only had a fixed gap size
between the cell walls. The open cell was constructed of ¼ in
float glass, separated by strips of Teflon of known thickness.  It
had dimensions comparable to the closed cell, but was open at the
top, so that beads could be dispersed through a funnel which
steadily dropped beads as it traveled back and forth across the
top of the cell (Fig.~\ref{fig2}).   In this way, the deposition
rate of the beads into the cell could be controlled precisely by
varying the speed and the size of the funnel. The present
experiments were designed to test the effect of changing the
distance between the walls on the roughness of the final interface
while keeping the interparticle distance constant.   The
experiments took place in the open cell and we investigated the
effect of variability in the gap by setting the gap at different
values, and measuring the effect of the walls on the roughness of
the final interface formed after sedimentation. We investigated
gaps ranging from 0.8 mm to 2.0 mm. The ratio of the gap thickness
to the bead diameter was defined as a dimensionless parameter $R$,
and our experiments spanned a gap/bead diameter ratio of $R =
1.33$ to $R = 3.33$. The previous experiments took place at either
a gap of 1mm ($R = 1.66$) (closed cell) or 0.8 mm ($R = 1.33$)
(open cell), so this range of gap values went far beyond our
earlier measurements. In all cases, the deposition rate of beads
into the cell was controlled at about 4 beads/sec so the average
distance between the particles was about $20R$. The surface was
photographed during and at the end of the deposition process and
the photographs were digitized by a Nikon LS-2000 film scanner.
Individual particles were typically resolvable and thus the
position of the particles on the interface could be traced
accurately. There is a limit to the extent over which the gap can
be widened without changing the method of analysis, since at one
point it will no longer be possible to analyze the rough surface
as a one-dimensional interface. We believe that we are already
past that limit at $R = 2$, but to give an estimate of the effects
of the wall separation to the interested reader we have included
the data for $R > 2$.

\section{DISCUSSION\protect\\ }
\label{sec:level3} As in the previous work, we have analyzed these
rough surfaces using the scaling ansatz proposed by Family and
Vicsek \cite{Family85}.   In this ansatz, the rms thickness of the
interface is defined to be:

\begin{equation}
W (L,t) = \left[ \frac{1}{N} \sum_{i=1}^N \tilde{h}(x_{i},t)^{2}
\right]^ {\frac{1}{2}} \label{eqn1}
\end{equation}
where
\begin{equation}
   \tilde{h}(x_{i},t) = h(x_{i},t)-\bar{h}(t)
\label{eqn2}
\end{equation}
and
\begin{equation}
   \bar{h}(t) = \frac{1}{N} \sum_{i=1}^{N} h(x_{i},t) .
\label{eqn3}
\end{equation}

The scaling ansatz predicts that:

\begin{equation}
   W(L,t)= L^{\alpha} f(t/L^{\alpha / \beta})
\label{eqn4}
\end{equation}

where the exponents $\alpha$ and $\beta$ are the static and
dynamic scaling exponents.   The function $f(t/L^{\alpha /
\beta})$ is expected to have an asymptotic form such that

\begin{equation}
   W(L,t) \sim t^{\beta} \; for \; t \ll L^{\alpha / \beta}
\label{eqn5}
\end{equation}
and
\begin{equation}
   W(L,t) \sim L^{\alpha} \; for \; t \gg L^{\alpha / \beta}
\label{eqn6}
\end{equation}

\begin{figure}[bp]
\includegraphics[width=8cm]{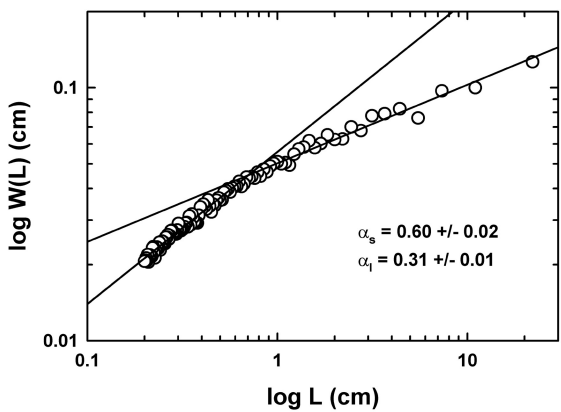}
\caption{Roughness function W(L,t) vs. L for a typical interface
at a gap of 0.8mm} \label{fig3}
\end{figure}

Fig.~\ref{fig3} shows an example of W(L,t) at a typical gap/bead
ratio. To minimize the wall effects at the horizontal edges, we
have used only the middle $70\%$ of each interface for our
analysis.

\begin{figure}[bp]
\includegraphics[width=8cm]{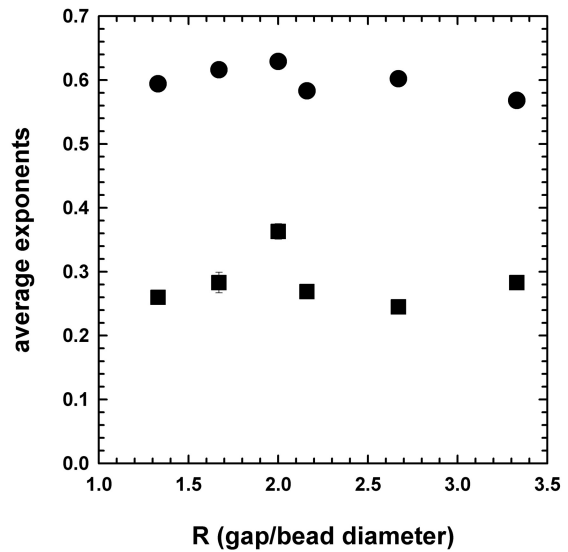}
\caption{Average roughness exponents found at different bead
diameter/gap ratios, R. The circles denote al, the squares
represent as. The values of each a are typically averaged over
four experimental runs, each of which would have had relatively
small uncertainty in a. Thus, the stated uncertainties arise from
the reproducibility of roughness.} \label{fig4}
\end{figure}

At all values of R (gap/bead ratio) studied, we still see two well
defined roughness exponents. These roughness exponents have a
crossover length scale at about 1 cm, which is typical from the
previous work. Our earlier work corresponded to a gap width of 1.0
mm, and the present work gives the same results at this gap width
as expected.   As the value of R is increased (Fig.~\ref{fig4}),
we do not see any significant change in the value of either
exponent. While there is a slight increase in the scaling
exponents around $R = 2$, this slight increase is less significant
than the increase in the roughness exponents when we increase the
mean particle separation by increasing the feeding rate of the
particles in our previous experiments, where the large length
scale roughness rose from ~0.2 at an average deposition rate of
one bead/sec to ~0.5 at an average deposition rate of 35
beads/sec.

We have investigated the effect of the interaction between the
walls of the container and the sedimenting particles on the
roughness exponent of the surface formed by this
quasi-two-dimensional sedimentation. The roughness exponent is
found to be robust to the changes in the separation between the
walls of the container. The reasons for the slight increase in the
of the roughness exponent at R = 2 is under further investigation
using computational methods to simulate the effects of the wall
separation, as well as direct experimental investigations of
correlations between particles in the fluid.

\begin{acknowledgments}
The authors would like to acknowledge the advice and loan of
equipment graciously provided by Dr. James Maher at the University
of Pittsburgh.
\end{acknowledgments}

\end{document}